\documentclass[aps,twocolumn,groupedaddress,superscriptaddress,floatfix,amsmath,amssymb,prb]{revtex4}
\usepackage{amsmath}
\usepackage{amssymb}
\usepackage{graphicx}
\usepackage{textcomp}
\usepackage{bm}

\usepackage{booktabs}
\usepackage{siunitx}

\usepackage[usenames,dvipsnames]{color}	

\begin{document}
\title{First-principles many-body models for \\
	electron transport through molecular nanomagnets}
    
\author{A. Chiesa}
\affiliation{Institute for Advanced Simulation, Forschungszentrum J\"ulich, 52425 J\"ulich, Germany}
\affiliation{Dipartimento di Scienze Matematiche, Fisiche e Informatiche, University of Parma, 43124 Parma, Italy}
\author{E. Macaluso}
\affiliation{Dipartimento di Scienze Matematiche, Fisiche e Informatiche, University of Parma, 43124 Parma, Italy}
\author{P. Santini}
\affiliation{Dipartimento di Scienze Matematiche, Fisiche e Informatiche, University of Parma, 43124 Parma, Italy}
\author{S. Carretta}
\email{stefano.carretta@unipr.it}  
\affiliation{Dipartimento di Scienze Matematiche, Fisiche e Informatiche, University of Parma, 43124 Parma, Italy}
\affiliation{UdR Parma, INSTM, I-43124 Parma, Italy}
\author{E. Pavarini}
\email{e.pavarini@fz-juelich.de}  
\affiliation{Institute for Advanced Simulation, Forschungszentrum J\"ulich, 52425 J\"ulich, Germany}
\affiliation{JARA High-Performance Computing,RWTH Aachen University, 52062 Aachen, Germany}

\date{\today }

\begin{abstract} 
Impressive advances in the field of molecular spintronics allow one to study electron transport through individual magnetic molecules embedded between metallic leads
in the purely quantum regime of single electron tunneling. Besides fundamental interest, this experimental setup, in which a single molecule is manipulated by electronic means, provides the elementary units of possible forthcoming technological applications, ranging from spin valves to transistors and qubits for quantum information processing.
Theoretically, while  for weakly-correlated molecular junctions established first-principles techniques do enable the system-specific 
description of transport phenomena,   
methods of similar power and flexibility are still lacking for junctions involving strongly-correlated molecular nanomagnets. 
Here we propose an efficient scheme  based on the {\em ab-initio} construction of material-specific Hubbard models
and on the master-equation formalism. 
We apply this approach to a representative case, the \{Ni$_2$\} molecular spin dimer, in the regime of weak molecule-electrodes coupling, the one relevant for quantum-information applications. 
Our approach allows us to study in a realistic setting 
many-body effects such as current suppression and negative differential conductance. 
We think that this method has the potential for becoming a very useful tool for describing transport phenomena in strongly correlated molecules.
\end{abstract}

\maketitle

\section{Introduction}
The emerging field of molecular spintronics has paved the way to the manipulation and read-out of
individual spins by electronic means, with an unprecedented degree of control.\cite{Thiele2014} The
contemporary exploitation of electronic and spin degrees of freedom at the single-molecule level can give rise to
a hybrid architecture, which combines the best characteristics of the two worlds: the fast, local electric
control \cite{Bogani,WW2012} and the protection from the detrimental effect of decoherence, ensured by the spins. \cite{Zadrozny2015,SEs} A possible application of this setup is provided by
quantum information processing (QIP).\cite{Grover17} Indeed, 
bottom-up nanofabrication techniques can be exploited to realize quantum computing architectures starting
from their individual components, namely from a set of interacting qubits. 
Potential building blocks for designing
such devices are molecular nanomagnets (MNMs), which can be used to encode qubits in QIP architectures\cite{Gatteschi,Pinkowicz,Luis2011,Luis2012,Aromi2014,SciRep14,Ardavan2015,NatComm16,Chem,VO2,Ybtrensal} potentially competitive with current leading  technologies. \cite{NatPhys2019}
MNMs are clusters
containing transition metal or rare-earth ions, embedded in an organic sheath that can be tailored to bind them
onto surfaces. The ability to control intra- and inter-molecular magnetic interactions almost at will by
coordination chemistry, thus realizing complex structures such as even- and odd-membered rings, \cite{RevRings,Cr9,Cr8Zn,Cr8Mn} and the remarkably long coherence times reported for some of them, \cite{Bader2014,Zadrozny2015,Tesi2015,Atzori2016,Atzori2016b,Freedman2016} 
makes these systems particularly attractive for technological applications. Electric read-out of the magnetization and even of the nuclear
spin state of single-molecule magnets has already been
demonstrated using a three-terminal geometry which acts as a single-electron transistor. \cite{Thiele2014,Urdampilleta2015} In this apparatus,
the MNM bridges the gap between the two conducting nanoleads and is also connected to a gate voltage, which
is used (in the regime of weak coupling to the leads) to control the quantized charge on the MNM. As soon as
bias or gate voltages lift Coulomb blockade, transport occurs via tunneling of single electrons in and out of the
molecule. Hence, information about the magnetic properties of the MNM can be obtained by transport
measurements.\cite{PRLFe42012,PRBFe42015}

In view of designing new platforms for QIP,  
{first-principles methods are essential tools}
to characterize the behavior of MNMs embedded in a molecular spintronics architecture. \cite{PedersonRev} 
Indeed, only these approaches can provide the
detailed understanding of the mechanisms underlying inter- and intra-molecular interactions, 
which is key to realize an efficient QIP scheme. 
{Unfortunately, typically MNMs are also strongly-correlated molecules,
and for strongly-correlated molecules the {\it ab-initio} description
of transport experiments remains to date a challenge. 

The most commonly adopted theoretical approaches to describe transport in molecular devices fall in two categories. The first category is the one of
{\em ab-initio} methods based on density-functional theory (DFT), typically combined with
either the Landauer-B\"uttiker method or non-equilibrium Green functions. In studies based on these approaches the material aspects
are successfully taken into account  (see, e.g., Refs. \onlinecite{Sanvito,Ryndykbook,PedersonFe4,Michalak}), but correlation effects are treated at a mean-field-like level, via simple approximations to the exact DFT exchange-correlation functional. 
However, neither LDA/GGA, LDA/GGA+$U$ nor hybrid functionals 
correctly describe the excited spectrum of a strongly-correlated molecule. This class of approaches  is thus bound to fail in properly capturing phenomena escaping the static mean-field description. Examples are the complete current suppression or the negative differential conductance.\cite{Cuniberti,JCPWegewijs2017}}
Alternative first-principles schemes are, e.g., based on time-dependent density-functional theory; the latter however treats transport  as a time-dependent phenomenon instead of focusing on steady-state properties. For steady-state properties, 
particularly promising appears the recently proposed
$i$-DFT scheme, in which the exchange-correlation potential depends non only on the density but also on the steady current. Unfortunately, however,  the $i$-DFT  exchange-correlation potentials is so far only known for simple  exactly solvable
many-body models.\cite{Kurth}

{The second category of approaches is based on effective 
models (see, e.g., Refs. \onlinecite{Park2016,PRLWegewijs2006,PRBTimm2006}).  
Here, non-trivial many-body effects beyond static mean-field are correctly described. The models are however typically empirical,  often based on low-energy spin-only Hamiltonians, and
their parameters
are usually obtained by fitting experiments.   This limits their actual predictive power, and makes it difficult to
account for non-trivial material  aspects.
So far, very few attempts to go beyond this, building many-body Hamiltonians {\it ab-initio},
were reported. Among these,  Ref.~\onlinecite{Richter} for a  junction involving the molecule S-C$_6$H$_4$-S and
Ref.~\onlinecite{Yu} for dicyanovinyl-substituted quinquethiophene, focusing on LUMO and LUMO+1 states.
For complex molecular systems with one or more ions with $d$ and/or $f$ open shells  to be included in the model, this remains a great challenge. }

In this work we propose an alternative practical scheme, applicable in principle to strongly-correlated molecules of any complexity,
and which allows us to treat both many-body and material aspects on the same footing.
The scheme is designed for the weak molecule-electrode coupling regime, in which the potential qubits keep the properties of the isolated molecules and hence are promising for QIP applications.
We show the power of the method for a prototypical case, the \{Ni$_2$\} molecule.\cite{PRBNi2,Dalton}
Transition metal dimers of this form have already been studied as test-beds for transport phenomena, such as Kondo effect or singlet-triplet switching by a bias voltage.\cite{NatNanoCo2}
For \{Ni$_2$\}-based junctions (see Fig.~\ref{fig1}), we predict clear signatures of strong correlation effects, and in particular the onset of spin blockade 
and negative differential conductance. 
Based on our system-specific model, we can relate these phenomena to the intrinsic properties of the system studied, and  determine the optimal experimental set up
for which they can be observed.

The proposed approach is an extension to transport through molecular junctions of the method we recently introduced in Ref.~\onlinecite{PRLdft} for describing the magnetic properties of correlated nanomagnets.
It combines density-functional theory (DFT) and many-body (MB) methods, and thus we refer to
it in short as DFT+MB approach.
We have already proved that this technique is very successful for the description of the magnetic properties of MNMs at equilibrium.\cite{PRLdft,SciRep14,PRBdft,ChemSci}
Our approach has no free parameters and does not rely on a phenomenological description of the molecule via a spin Hamiltonian, which only holds when charge fluctuations are negligible.
For describing transport experiments we proceed as follows. First we build system-specific Hubbard models for the molecule+leads system. In this initial step we use the model-building part of the DFT+MB  approach as introduced
in Ref.~\onlinecite{PRLdft}. Next we use the Hubbard Hamiltonians obtained in this way to set up system-specific
master equations. The solution of the latter allows us 
to calculate the stability diagrams typically measured in transport experiments. In this second step, it would be in principle necessary to diagonalize
the full Hubbard model in Fock space. For the \{Ni$_2$\} molecule this can be done exactly without exploiting symmetries.
In general, for large molecules, the size of the Fock space becomes quickly prohibitively large, however. 
This is the bottleneck of the approach.
Nevertheless, here we show that one can use the irreducible tensors technique, extended to {\em fermionic} operators,\cite{ITOs,ITOs2,Biedenharn} to strongly decrease both memory needs and computational time.
This 
irreducible-tensors-based approach extends the perspective application of our scheme to significantly larger molecules. \\
The paper is organized as follows.
In Section II we introduce the {\it ab-initio} DFT+MB approach to electron transport through correlated molecular junctions. 
Here we also explain how we set up and solve the master equations and the fermionic irreducible tensor technique.
In Section III we present the results for the prototypical spin dimer Ni$_2$, with focus on the peculiar many-body signatures emerging from our description. We finally draw the conclusions in Section IV.

\section{Methods}
\begin{figure}[b]
	\centering
	\includegraphics[width=0.54\textwidth]{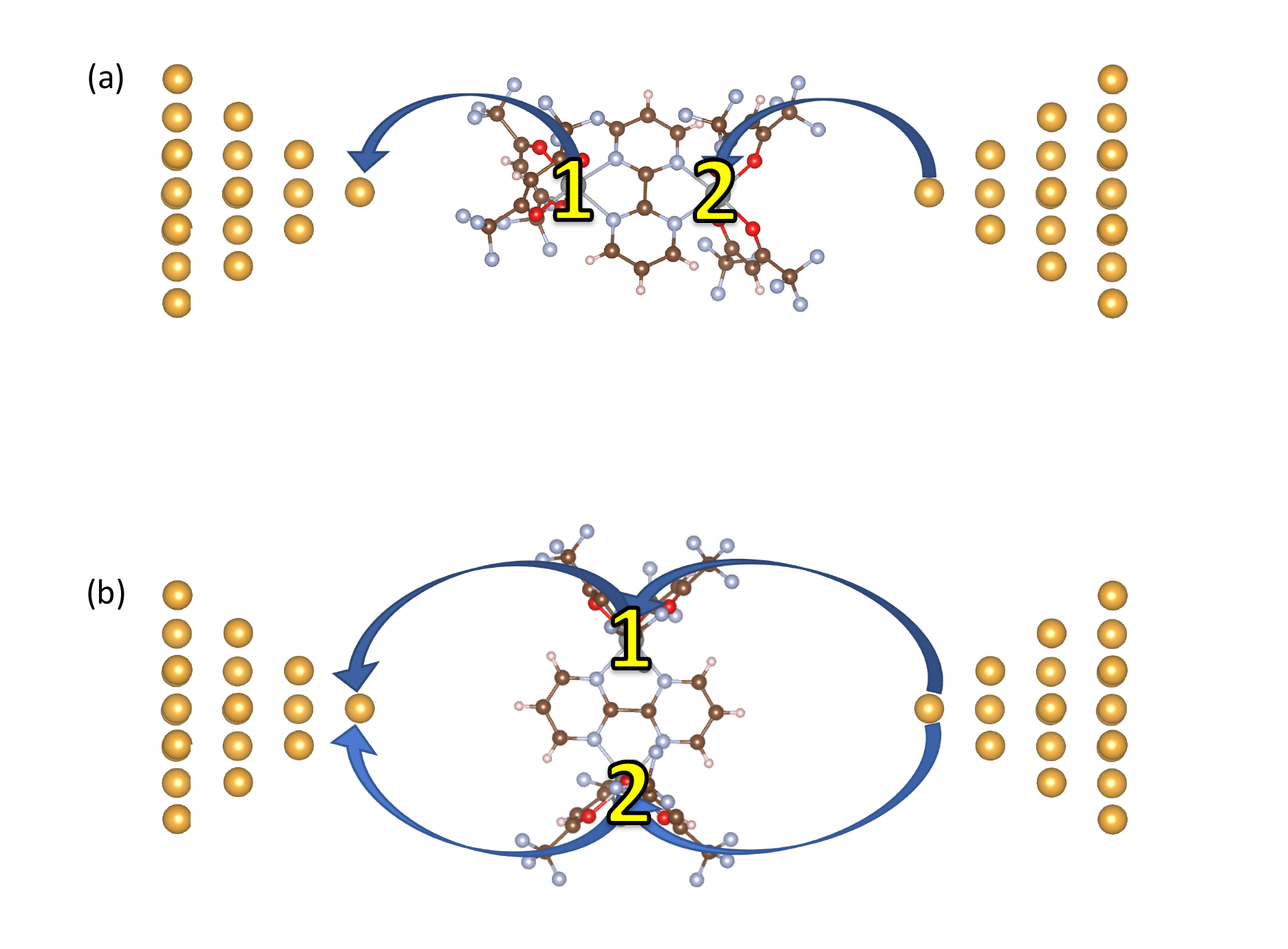} 
	\caption{Structure of the molecular junction in two configurations: serial (a) and parallel (b).
		The \{Ni$_2$\} molecule is placed at a $\sim 6-10~$\AA\, distance from the Au clusters.}
	\label{fig1}
\end{figure}

In this work, we focus on the weak molecule-electrode coupling regime,
where the molecular properties remain almost unaltered by the contact with the electrodes.
In this regime the molecule is sufficiently far
from the metallic leads, the hybridization is weak and
the specifics of the electrodes are not so important. 
This allows us to concentrate on molecular properties (rather than on the molecule-electrodes coupling, which is specific of any experimental implementation)
and hence to use the leads only as a manipulation tool of the molecular state. This is also the most interesting regime for QIP applications, \cite{Grover17} reducing decoherence originating from the coupling to the metallic electrodes.
Our procedure can be then summarized in these three steps:
\begin{enumerate}
	\item We first perform DFT+MB calculations \cite{PRLdft,SciRep14,PRBdft,ChemSci} for the system consisting of the target  molecule embedded between  two gold clusters (see Fig.~\ref{fig1}).  Using this approach, we extract the parameters of the generalized Hubbard model for the molecule and the associated molecule-leads tunneling rates. 
	\item Next we diagonalize the Hubbard model (in the zero molecule-leads tunneling limit) and obtain the {\it molecular} many-body states for each charge sector, with fixed number of electrons $N$. This is done by exploiting molecular point-group simmetries and rotational invariance (in the limit of weak spin-orbit coupling) and with the help of the fermionic irreducible tensor operators method.
	\item We then  write the master equation for the population of the molecular Fock states and look for the steady-state solution.
	 We finally compute observables { -- here the current and differential conductance --} as a function of bias and gate voltage.
\end{enumerate}
{In the next subsections we give additional details on each of the three steps.}

\subsection{DFT+MB approach}
We construct system-specific generalized Hubbard models along the lines of Refs.~\onlinecite{PRLdft} and \onlinecite{PRBdft}. With respect our previous works, the main difference in this step is that the system under investigation (Fig.~\ref{fig1}) consists of a {\it correlated}-electron molecule embedded between two metallic electrodes containing {\it uncorrelated} electrons. 
First, we perform DFT calculations in the local density approximation (LDA) for the whole system. The electrodes are  described by chemically stable finite Au clusters (see below), which can be treated as weakly correlated. \cite{geometry}
Calculations are based on the NWChem code, \cite{nwchem} and we employ a triple-zeta valence basis set to describe the molecule. 
The core Au orbitals are included into the effective (pseudo) potential LANL2DZ.\cite{Herrmann} 

In the second step, we identify the transition metal $d-$like states around the Fermi level and apply the Foster-Boys localization procedure.\cite{Boys} 
The $d$-like Foster-Boys orbitals obtained in this way span the states close to the Fermi level;
 the coupling to the ligands via hybridization is accounted for by construction, as can be seen from the tails of the orbitals on the ligands.
This Foster-Boys basis is then used to build a generalized Hubbard model, consisting of three terms:
\begin{equation}
H = H_{mol}+H_{el}+H_T.
\label{Htot}
\end{equation}
The first term is the {\it correlated} molecular Hamiltonian, and has the form: 
\begin{eqnarray}\nonumber
H_{mol}&=& - V_G  N   -\sum_{ii^\prime\sigma}\sum_{mm^\prime} t^{i,i^\prime}_{m,m^\prime} c^\dagger_{im\sigma} c_{i^\prime m^\prime\sigma} \\ \nonumber
&+&\frac{1}{2} \sum_{i i^\prime\sigma\sigma^\prime}\sum_{mm^\prime}\sum_{p  p^\prime}
U^{i,i^\prime}_{m p m^\prime p^\prime}
c^\dagger_{im\sigma} c^\dagger_{i^\prime p\sigma^\prime} c^{\phantom{\dagger}}_{i p^\prime\sigma^\prime}c^{\phantom{\dagger}}_{i^\prime m^\prime\sigma} \\ 
&+&\sum_i \lambda_i\; \sum_{mm'\sigma\sigma'} \xi^i_{m\sigma,m'\sigma'}c^\dagger_{im\sigma} c_{i m^\prime\sigma'} -H_{\rm DC}.
\label{hubbard}
\end{eqnarray}
Here $c_{im\sigma}^\dagger$ ($c_{im\sigma}^{\phantom{\dagger}}$) creates (annihilates) a $3d$ electron on the molecule with spin $\sigma$ in the Boys orbital $m$ at site $i$. \\
$N = \sum_{im\sigma} c^\dagger_{im\sigma} c_{i m\sigma}$ is the total number operator and $V_G$ indicates the gate potential energy, acting as a chemical potential on each molecular orbital. 

The parameters $-t^{i,i^\prime}_{m,m^\prime}$ are the hopping ($i\ne i^\prime$) or the crystal-field ($i=i^\prime$) integrals. In the following we indicate the energy of the crystal-field orbitals (obtained by diagonalizing the the on-site matrix $t^{i,i}_{m,m^\prime}$) with $\varepsilon_m$ and order them such that $\varepsilon_m \le\varepsilon_{m+1}$. 
Since the Ni$^{2+}$ ions are in a octahedral-like environment, the levels are approximatively
split into lower energy $t_{2g}$-like states and higher energy $e_g$-like states, so that
the ionic ground configuration can be described as $t_{2g}^6e_g^2$, with total spin $s_i=1$.

The terms $U^{i,i^\prime}_{m p m^\prime p^\prime}$ are the screened Coulomb integrals.
For simplicity here we use the rotationally-invariant Kanamori form of the Coulomb vertex.
In this approximation all on-site Coulomb parameters can be expressed as a function of the averaged screened Coulomb couplings $U^{i,i}=U$ and $J^{i,i}=J$, which, in turn, depend only on the Slater integrals $F_0$, $F_2$ and $F_4$.\cite{book} The essential Coulomb integrals are the direct ($U_{mm'mm'}^{i,i} =U_{m,m'}=U-2J(1-\delta_{m,m'})$) and the exchange ($U_{mm'm'm}^{i,i}=J$) interaction, the pair-hopping ($U_{mmm'm'}^{i,i}=J$) and the spin-flip term ($U_{mm'm'm}^{i,i}=J$).
We calculate $U$ and $J$ via the constrained LDA (cLDA) \cite{cLDA} approach in the Foster-Boys basis,
keeping the basis frozen in the self-consistency loop.
We find $U=6.3$ eV and $J=0.26$ eV for the Ni$_2$ junction shown in Fig.~\ref{fig1}.
In addition, nearest-neighbors Coulomb exchange integrals are evaluated by cLDA (see below).
$H_{\rm DC}$ is the double-counting correction,
which removes the mean-field part of the local Coulomb interaction, already included in the LDA.
Here we adopt the fully localized limit.\cite{book} 
Finally, $\lambda_i$ is the strength of the spin-orbit interaction, here  the same for all the $3d$ electrons within the same ion. The terms $\xi^i_{m\sigma,m'\sigma'}=\langle m\sigma |{\bf s}^i\cdot\boldsymbol{\ell}^i |m'\sigma'\rangle$ are matrix elements
of the spin-orbit matrix in the Forster-Boys basis. In our approach,
$\lambda_i$ can be extracted by comparing the single-electron crystal-field splittings with and without spin-orbit interaction.
We have already shown \cite{PRLdft} that for Ni$^{2+}$ in an octahedral environment (as in the system studied in the present work)  a very good approximation of $\lambda_i$ can be obtained by using tabulated single-ion values.\cite{Abragam} We thus here adopt this strategy, to avoid time-consuming relativistic self-consistent calculations. 

The second term in Eq. (\ref{Htot}) is 
$$
H_{el} = \sum_{l=L,R} \sum_{k,\sigma} \epsilon_{l k\sigma} a^\dagger_{l k\sigma} a_{l k\sigma},
$$ and models the {\it uncorrelated} left (L) and right (R) electrodes. Here $a^\dagger_{l k\sigma}$ ($a_{l k\sigma}$) creates (destroys) an electron with energy $\epsilon_{l k\sigma}$ on orbital $k$ of the electrode $l$. The energies $\epsilon_{l k\sigma}$ are obtained by diagonalizing the part of the one-electron Hamiltonian on each of the two clusters representing the electrodes. \\
Finally, the tunneling Hamiltonian $$H_T = \sum_{im} \sum_{lk\sigma} \tau^{li}_{km} a^\dagger_{lk\sigma} c_{im\sigma}+{\rm h.c.}$$ describes the tunneling processes between the two electrodes and each $d$-like molecular orbital.

For the \{Ni$_2$\} junction, one might wonder, if, in addition to Ni-centered $d$-like Foster-Boys orbitals, the $p$ orbitals of the pyridine in between the two Ni centers play a role and should be explicitly accounted for. We find
that indeed one of these orbitals is close to the Fermi level and is coupled to both Ni ions.
The associated hopping integrals are however significantly smaller than the energy gap (the ratio is $\sim 1/4$). 
We have checked by including this $p$ orbital in the generalized Hubbard Hamiltonian (\ref{hubbard}) and performing full diagonalization that many-body states are only slightly modified, with a negligible occupation of the $p$ orbital in the anion.  Hence, for simplicity, we neglect this orbital in the discussion that follows.

\subsection{Master equation description of transport}
In the weak-coupling limit between the electrodes and the molecule,\cite{Richter,Park2016} $H_T$ acts as small perturbation of the non-interacting Hamiltonian $H_{el}+H_{mol}$. In the absence of molecule-lead coupling ($H_T=0$), we can 
build the many-body states of the whole system as 
a tensor product of the separate eigenstates of $H_{el}$ and $H_{mol}$.
In the following we label as $|\lambda_N \rangle$ the eigenstates of $H_{mol}$ 
with energy $E_\lambda$ and $N$ electrons. 
As we will discuss later in more detail, the spectrum of the isolated \{Ni$_2$\} molecule consists of a sequence of total spin multiplets, split by the spin-orbit interaction. 
These spin-orbit-induced splittings ($\delta E$) are larger than the molecule-leads tunneling rates (we find $\delta E\sim 3-6$ K) and the leads
are non magnetic. Therefore interference effects, which have been shown to be very important in the presence of degenerate or almost degenerate states,\cite{Beenakker,Braun,Schultz,Begemann,Donarini2009,Donarini2019} are strongly suppressed.  
Thus we can separate the dynamics of diagonal and off-diagonal elements of the system density matrix. In the case of the \{Ni$_2$\}  junction considered here, current and conductance are therefore already accurately described from the stationary solution of the Pauli master equation \cite{vanKampen}
\begin{equation}%
\frac{dP_{\lambda_{N+1}}}{dt} = \sum_{\lambda^{\prime}} \left( R^{\lambda\lambda^\prime} P_{\lambda^\prime_N}-R^{\lambda^\prime\lambda} P_{\lambda_{N+1}} \right),
\label{masterEq}
\end{equation}
where $P_{\lambda^\prime_N}$ is the occupation probability of state $\vert \lambda^\prime_N \rangle$ and $R^{\lambda \lambda^\prime}$ is the rate matrix, representing the tunneling probability from the initial state $\vert \lambda^\prime_N \rangle$ to the final state 
$\vert \lambda_{N+1} \rangle$. To second order in $H_T$, this is given by:
\begin{align} \label{Rate} %
R^{\lambda\lambda^\prime} &= \sum_l \{ \gamma_{\lambda\lambda^\prime}^l f (\Delta^l_{\lambda,\lambda'}) 
+ \gamma_{\lambda^\prime\lambda}^l [1-f(\Delta^l_{\lambda',\lambda})]\}.
\end{align}
Here  
$f(E)$ is the Fermi-Dirac distribution function, $\Delta^l_{\lambda,\lambda'}=E_\lambda - E_{\lambda^\prime} - \mu_l$, 
and $\mu_l$ is the chemical potential of electrode $l$. Furthermore, 
\begin{eqnarray}\nonumber
\gamma_{\lambda\lambda^\prime}^l &=& \frac{2\pi}{\hbar} \sum_{\substack{ i i^\prime \\ m m^\prime}} \sum_{k \sigma} \tau^{li*}_{km} \tau^{li^\prime}_{km^\prime} 
\langle \lambda_N^\prime \vert c_{im\sigma} \vert \lambda_{N+1}  \rangle \\ 
&& \langle \lambda_{N+1} \vert c^\dagger_{i^\prime m^\prime \sigma} \vert \lambda_{N}^\prime \rangle \delta \left( E_\lambda - E_{\lambda^\prime} -\epsilon_{l k\sigma} \right).  
\label{gamma}
\end{eqnarray}
For Au electrodes 
the density of states $\rho(\epsilon)$ 
and the imaginary part of the hybridization function $\Gamma^l_{imi^\prime m^\prime} (\epsilon) = 2\pi \sum_{k\sigma}  \tau^{li*}_{km} \tau^{li^\prime}_{km^\prime} \delta(\epsilon-\epsilon_{l k\sigma})$
can be considered approximatively flat close to the Fermi level (wide-band limit).\cite{Herrmann,Richter} 
The specific value of $\rho(0)$  then merely yields a rescaling of the current and is thus irrelevant to describe specific molecular transport features. 
Here we model the electrodes (which are used to compute the coefficients $\gamma_{\lambda\lambda^\prime}^l$) by means of tetrahedral Au$_{20}$ clusters, which are chemically stable and have a large HOMO-LUMO gap of $\sim 2$ eV; it has been already established in the past that such clusters provide a good approximation of the metallic junction, \cite{Au20,PedersonFe4}
and already show all essential characteristics of the bulk band of gold.\cite{Hakkinen}  

In order to find the steady-state solution to Eq. (\ref{masterEq}), we solve $dP_\lambda/dt=0$ via the biconjugate gradient stabilizer algorithm with zero-bias Boltzmann distribution as initial occupation probabilities.\cite{Park2016}
Then, the current from electrode $l$ to the molecule is obtained as $I_l=I_l^{\rm i}+I^{\rm o}_l$, where
the two contributions account for electron hopping into ($I_l^{\rm i}$) and out ($I_l^{\rm o}$) of the molecule,
and
\begin{align}\nonumber
I^{\rm i}_l &=+e \sum_{\lambda \lambda^\prime} \gamma^l_{\lambda \lambda^\prime} f ( \Delta^l_{\lambda,\lambda^\prime}) P_{\lambda^\prime} \\
I^{\rm o}_l&=- e \sum_{\lambda \lambda^\prime} \gamma^l_{\lambda \lambda^\prime} \left[ 1 - f ( \Delta^l_{\lambda,\lambda^\prime})\right] P_{\lambda} .
\label{current}
\end{align}
In the stationary limit \cite{Cuniberti}, $I_R = -I_L$ and the total current is simply $I = \left(I_R-I_L\right)/2 = I_R$.
In the following, we compute current $I$ and the differential conductance $dI/dV$ as a function of bias and gate voltage,
assuming that the bias voltage is symmetric ($\mu_{L,R} = \pm V$). %

As previously discussed, Eqs. (\ref{Rate}-\ref{gamma}) are the results of a perturbative expansion to the lowest order in
$H_T$ (weak-coupling limit), in which the small parameter is $\Gamma/k_BT$ ($\Gamma = {\rm max} ~\Gamma_{imi^\prime m^\prime}$). %
Within this description only single-electron tunneling processes are relevant. As a rule of thumb, the single-electron tunneling condition is fulfilled if the typical time between two tunneling events is much larger than the time required to thermalize the excitations created in the metallic reservoirs.\cite{Cuniberti}
If the tunneling rate $\Gamma$ becomes larger than $k_BT$, co-tunneling events will in general take place. These can be accounted for by going to higher orders in the perturbative expansion.\cite{JCPWegewijs2017}
Still, the most important non-equilibrium effects %
are already captured correctly at the second
order, which  is often sufficient\cite{PRBFe42015} to describe well experimental results  even for $\Gamma \sim k_B T$. %
For our specific case, assuming typical values of $\rho(0)$, one can estimate $\Gamma \lesssim$ 1 K. 
Most of the transport spectra that will be discussed in the next sections of the paper are computed at $T=2$ K, a typical temperature of many experimental settings, see, e.g., Ref. \onlinecite{PRBFe42015}. In this regime, only the ground state of \{Ni$_2$\} is populated and  the ratio $\Gamma/k_BT$ remains sufficiently small to make co-tunneling events negligible. In these conditions, the essential features of the transport dynamics are indeed already 
well captured by lowest-order processes in $\Gamma$.\cite{PRBFe42015} In Section III we show that in the single-electron tunneling limit (where co-tunneling is neglected) the effect
of a finite temperature $T$ is merely to smooth out current features in the stability diagram. 

\subsection{Fermionic irreducible tensor operators}
For molecules containing several transition metal ions, the exact diagonalization of the generalized Hubbard model 
$H_{\rm mol}$ is a particularly hard task. Indeed, the Fock space of Hamiltonian (\ref{hubbard}) grows very quickly with the number
of orbitals and sites, thus making it impossible even to find the lowest eigenvalues and eigenvectors with the Lanczos method.
To minimize the size of the Hamiltonian blocks to diagonalize, we use symmetries.
First we exploit the conservation of the number of electrons $N$,
 $[H_{mol},N]=0$, and, in the absence of spin-orbit interaction, the conservation of the total spin $S$,
  $[H_{mol},{\bf S}]=0$.
We then rearrange the Hamiltonian in ($N,S$) blocks, decoupled from each other for $\lambda_i=0$.
Point symmetries can also be used at this stage, if present.
The core problem is the calculation of the many matrix elements due to the inter-site hopping term of the Hamiltonian. 
To minimize the numerical effort it is key to identify which Hamiltonian blocks are essential to calculate.
We do this by
recasting the latter in the form of compound irreducible tensor operators $T^k_q$ of rank $k=0$, obtained as the product of two rank $k_i=1/2$ tensor operators corresponding to the fermionic creator and annihilator.\cite{ITOs} More specifically, 
in the case studied here we have
\begin{align}\nonumber
\sqrt2 T^{0}_{0}(i m,  i^\prime m^\prime) &=   V^{1/2}_{1/2}(im) U^{1/2}_{-1/2}(i^\prime m^\prime)  \\ \nonumber 
&
- V^{1/2}_{-1/2}(im) U^{1/2}_{1/2}(i^\prime m^\prime)  
\label{ITO}
\end{align}
where the fermionic operators are 
$
V^{1/2}_{\sigma}(im)= c^\dagger_{im\sigma} $ and $
U^{1/2}_{\sigma}(im) = (-1)^{1/2-\sigma} c_{im -\sigma} $. Thus
the hopping term of Hamiltonian (\ref{hubbard}) takes then form 
\begin{align*}
\!\!\sum_{i\ne i^\prime}\sum_{\sigma mm^\prime} \!\! t^{i,i^\prime}_{m,m^\prime} c^\dagger_{im\sigma} c_{i^\prime m^\prime\sigma}=-\sqrt{2} \sum_{i\ne i^\prime m m^\prime} t^{i,i'}_{m,m'} T^{0}_{0} ( i m,  i^\prime m^\prime ).\end{align*} 
The matrix elements of the scalar operator $T^0_0$ in the total spin basis do not depend on the value of $S_z$. This greatly reduces the number of matrix elements that need to be computed. For a dimer we have
\begin{eqnarray} \nonumber
 && \langle \alpha_1 s_1 \alpha_2 s_2 S \vert T^0_0 \vert \alpha_1^\prime s_1^\prime \alpha_2^\prime s_2^\prime S^\prime \rangle =  \\ \nonumber
 && -\delta_{SS^\prime}\sqrt{2(2S+1)} 
   \begin{Bmatrix} 
0  & S & S \\
1/2  & s_1 & s_1^\prime \\
1/2  & s_2 & s_2^\prime
\end{Bmatrix}  \\ \nonumber
&& \langle \alpha_1 s_1 || V^{1/2} (1m) || \alpha_1^\prime s_1^\prime \rangle
\langle \alpha_2 s_2 || U^{1/2} (2m^\prime) || \alpha_2^\prime s_2^\prime \rangle.
\end{eqnarray}
Here all relevant ionic spin multiplets $s_i$ are included for each ion $i=1,2$, and they are labeled by the additional quantum number $\alpha_i$.
To derive the formula above, we used the Wigner-Eckart theorem  in conjunction with the recoupling technique, which allow us  
to write the matrix element above as the product of a 9$j$ symbol and two reduced single-site matrix elements, $ \langle \alpha_i s_i || V^{1/2} (im) || \alpha_i^\prime s_i^\prime \rangle$; the latter are by construction independent from the third component of $s_i$. 
The procedure can be generalized to a system consisting of several ions, taking care of the order of anti-commuting fermionic operators acting on different sites in the recoupling scheme.
Finally, the computed Hamiltonian matrix, now including inter-site hopping terms and on-site energies, is diagonalized in the separate ($N,S$) sectors. 
Since the spin-orbit interaction is small in $3d$ systems,\cite{Abragam,PRLdft,PRBdft} we 
treat it afterwards in second-order perturbation theory in the ($N,S$) basis described above. \\

\section{Results for the $\{$N\MakeLowercase{i}$_2$$\}$molecular nanomagnet}

\subsection {Model and low-energy many-body states}
\begin{figure}[t]
	\centering
	\includegraphics[width=0.45\textwidth]{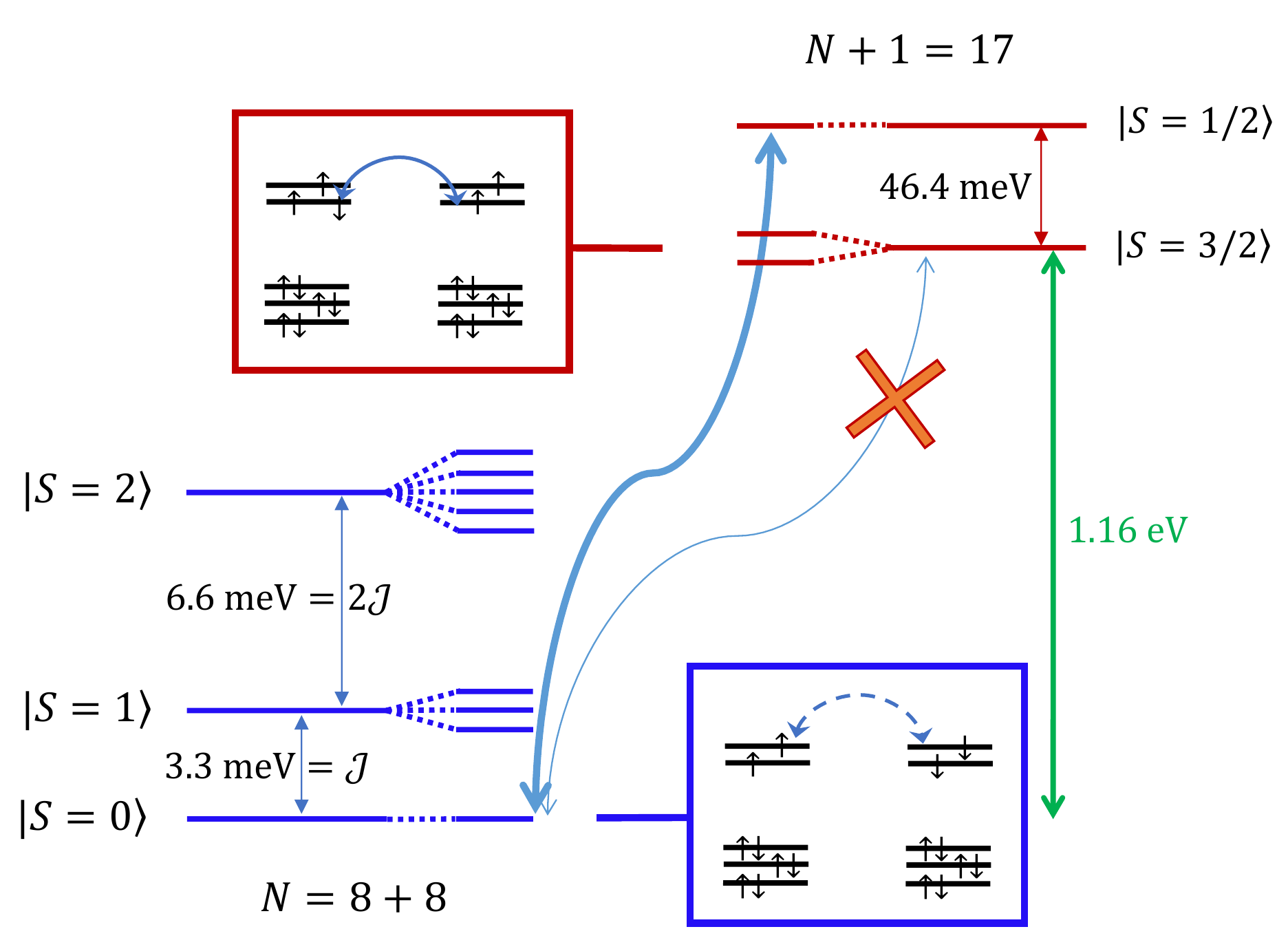} 
	\caption{(Color on-line) Schematic level-diagram of Ni$_2$ showing the lower energy  molecular multiplets, for $N=8+8$ and $N+1$ electrons, split by spin-orbit interaction.
		Neutral molecule ($N$ electrons, left): the lower-energy multiplets arise from %
		$s_i =1$ ionic configurations. States from $s_i \ne1$ configurations are  $\sim 0.5$ eV higher and can be neglected.
		Anion ($N+1$ electrons on the molecule, right): the ground quartet (Nagaoka state) and the first excited doublet are shown; the next excited states are
		two  doublets and a quartet,  and have energies  $ 135$ meV or higher above the ground state.
		Cation states ($N-1$ electrons on the molecule) are all very high in energy ($\sim 3$ eV) and are not shown.
		For each case, examples of the relevant ground ionic configurations are shown in the insects;  virtual  ($N$) and real ($N+1$) hopping processes are depicted with vectors. The minimal charge transfer energy  is $E(N+1,3/2)-E(N,0)\sim 1.16$ eV. In the absence of spin-orbit interaction, the transition between the ground states in the sector with with $N$ and $N+1$ electrons is forbidden, since $\Delta S>1/2$ (spin blockade). The transition probability is finite but small when the spin-orbit interaction is included; this is shown in the figure via a thin arrow.} 
	\label{fig2}
\end{figure}

We model the molecular junction as shown in Fig. \ref{fig1}. We consider two different coupling geometries: a serial configuration [panel (a)], in which each Ni ion is coupled to a single lead, and a parallel one [panel (b)], in which both Ni ions are coupled to both electrodes. \\
We first describe the molecular many-body states obtained by diagonalizing $H_{mol}$. The lowest eigenvectors belonging to the two charge sectors with $N$ and $N+1$ electrons are schematically depicted in Fig.~\ref{fig2}. They are separated by a charge transfer energy of 1.16 eV. States with $N-1$ electrons are much higher in energy ($\sim 3$ eV) and are not shown. 
Let us start considering the neutral molecule ($N=16$ electrons).
The lowest states for $N=16$ electrons arise from the ionic $t_{2g}^6e_g^2$ configurations with $s_i=1$ and are, respectively, a singlet, a triplet and a quintet. The effective spin-Hamiltonian describing this low-energy subspace is 
\begin{equation}
H_{\rm eff} = \mathcal{J} \textbf{s}_1 \cdot \textbf{s}_2 +\sum_i \textbf{s}_i \cdot \textbf{D}_i \cdot \textbf{s}_i.
\label{spinHam}
\end{equation}
The isotropic coupling $\mathcal{J}$ is the sum of the ferromagnetic Coulomb term (determined by cLDA) and the (here antiferromagnetic)
super-exchange coupling. 
Using the approach of Ref.~\onlinecite{PRLdft}, we find $\mathcal{J}=3.3$ meV, antiferromagnetic, and in good agreement with results from a recent inelastic neutron scattering study.\cite{Dalton}  
The zero-field splitting tensor $\textbf{D}_i$  is a full $3\times3$ matrix.
By diagonalizing it we can determine the principal anisotropy axes and the values of the axial $d_i$ and rhombic $e_i$ zero-field splitting parameters.
We find that both ions %
display easy-plane anisotropy, with the $z$ axis perpendicular to the plane of Fig. \ref{fig1}. We find $d_{1} = 90 ~\mu$eV, $d_{2} = 34 ~\mu$eV, $e_i \approx 0.12~ d_i$.
We also checked that anisotropic, as well as anti-symmetric contributions to the exchange interaction are negligible ($\lesssim \mathcal{J}/100$) in the present case. 
The three lowest total-spin multiplets (shown in the left part of Fig. \ref{fig2}) are separated by energies $\mathcal{J}$ and $2\mathcal{J}$ and are further slightly split by anisotropy. Excited molecular states originating from single-ion configurations with $s_i \neq 1$ are at least 500 meV above and are thus not shown in the schematic diagram
of Fig.~\ref{fig2}. 
\\
Let us now consider the case of the molecule with $N+1$ electrons, i.e., the anion.
Remarkably, in this case the ground multiplet is  a ferromagnetic  Nagaoka \cite{Nagaoka} state with maximum spin $S=3/2$. 
The reason is the following.
In the neutral molecule case, each ion is in the $t_{2g}^6e_g^2$ configuration, and magnetism is controlled by  super-exchange, which for \{Ni$_2$\}, as discussed above, is antiferromagnetic. 
If we add an extra electron to the molecule, however, super-exchange  is not the only possible origin of magnetic ordering.
As a matter of facts,  if the two ions are arranged ferromagnetically,
the extra electron can gain kinetic energy jumping between the two sites, without any Coulomb energy cost. This yields a first order energy gain in the associated hopping integral $t$, which dominates over 
super-exchange when the ratio $t/U$ is small, as in the cases considered here.
Nagaoka states were found to play an important role in the transport properties of other $d$ molecular complexes.
\cite{jacsTallal,Cuniberti,PRBWegewijs2007}
Remarkably, this has important consequences for transport.
Indeed, switching from a  total spin $S=0$ in the neutral molecule to {the maximum} spin $S$ in the charged (anion) molecule
gives rise to peculiar behaviors. In particular, transitions between the two lowest energy multiplets of adjacent charge sectors 
with a difference in total spins $\Delta S>1/2$
are forbidden in the absence of spin-obit coupling, leading to negative differential conductance and spin blockade\cite{JCPWegewijs2017} in some regions of the $(V,V_G)$ parameter space.
We will discuss this for \{Ni$_2$\} in the next subsections. \\

\subsection{Transport spectroscopy}
\begin{figure}[t!]
	\centering
	\includegraphics[width=0.4\textwidth]{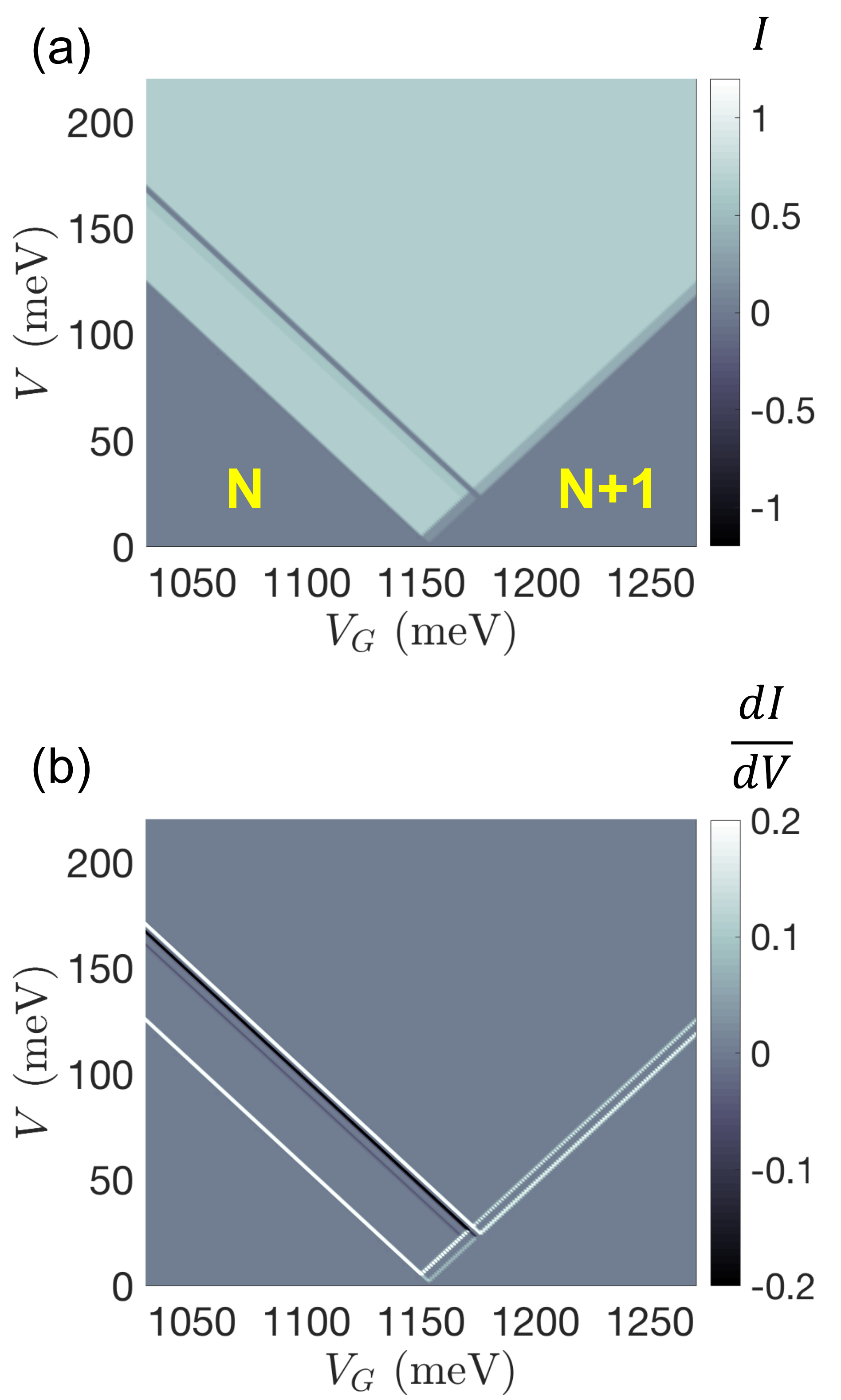} 
	\caption{(Color on-line) Calculated current $I$ (a) and differential conductance $dI/dV$ (b) as a function of bias ($V$) and gate ($V_G$) voltage at $T=2$ K. }
	\label{fig3}
\end{figure}

We first consider the molecular junction in Fig.~\ref{fig1}-a. In this configuration each Ni ion
is connected only to one electrode (serial configuration).
Fig. \ref{fig3} shows the calculated current and differential conductance  as a function of applied bias and gate voltages at $T=0$.
In the upper panel of the figure the current map $I(V,V_G)$ is shown.
Here the two areas labeled by $N$ and $N+1$ correspond to regions where transport is blocked ({\it Coulomb-blockade diamonds}) and the number of electrons on the molecule is thus either $N$ (left side) or $N+1$ (right side).\cite{RevvanderZant2015} By changing the bias or gate voltage, the blockade is lifted. Single-electron tunneling 
occurs when
the chemical potential of the molecule equals the Fermi energy of one of the electrodes; this is
what happens in the lighter and darker areas of Fig.~\ref{fig3}-(a). 
The lower panel of Fig.~\ref{fig3}, panel (b), shows the differential conductance map. Here the edges of the diamond of the upper panel become bright lines, corresponding to resonances. 

\begin{figure}[b!]
	\centering
	\includegraphics[width=0.45\textwidth]{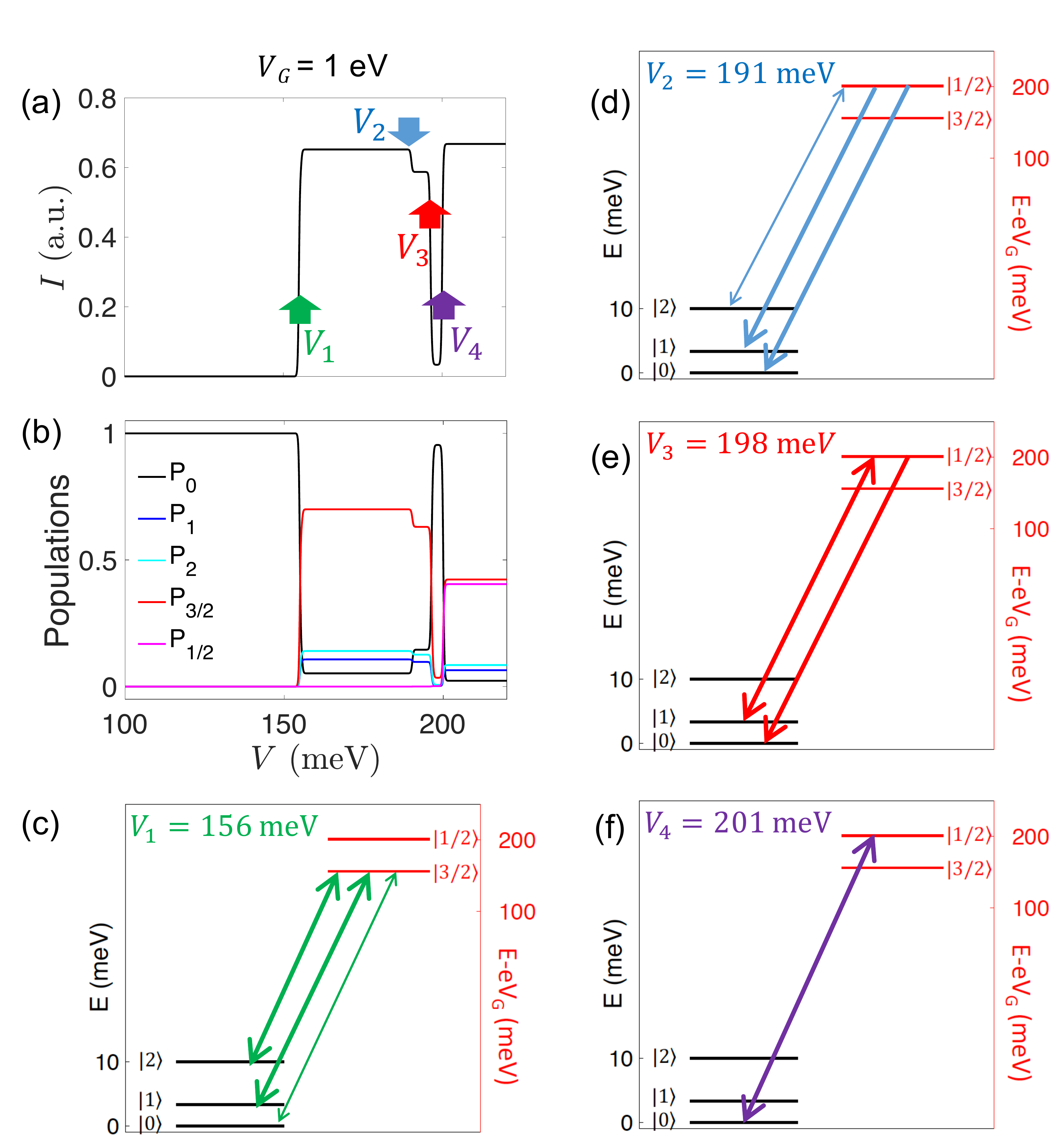} 
	\caption{(Color on-line) (a) Calculated $I(V,V_G)$ as a function of $V$ for fixed $V_G = 1$ eV. The labels $V_i$ ($i=1,\dots,4$), with $V_{i+1}>V_i$, indicate the positions of peaks in the current derivative $dI/dV$. (b) Corresponding populations $P_S$ of the relevant multiplets, labeled by their total spin $S$; the latter is integer for $N$ electrons and half-integer for $N+1$ electrons. 
		(c-f) Level diagram of the  spin multiplets $|S\rangle$ relevant at $V=V_i$. Left: $N$-electron states.
		Right: $N+1$-electron states.  Arrows indicate the associated transitions; 	the thickness is approximatively proportional to the actual transition probability. Very thin arrow indicate transitions which are forbidden
		in the absence of spin-orbit interaction. Double arrows mean
		that the transition is possible in both directions. }
	\label{fig4}
\end{figure}

In Fig.~\ref{fig3}-(b) there are additional resonance lines, parallel to those corresponding to the diamond edges.
These are electronic excitations.
In particular,  the dark line indicates  a narrow region of {\em negative} differential conductance.
A deeper insight into this peculiar feature can be gained by analyzing the conductance 
as a function of $V$ but with fixed $V_G$, and comparing it with the level diagrams of the two examined charge sectors.
This is done in Fig.~\ref{fig4}. Panel (a) shows $I(V,V_G)$ for $V_G=1$ eV. Here we explicitly indicate
with $V_i$ ($i=1,\dots,4)$ the four values of V which yield a sharp peak in the differential conductance.
Panel (b) shows the stationary population of the corresponding key many-body multiplets with $N$ and $N+1$ electrons. 
Panels (c-f) show instead, for each $V_i$, the energy levels of the neutral (black lines, left) and of the charged molecule (light lines, right), and the associated multiplets $|S\rangle$. Spin-orbit effects, among which the zero-field splitting, are included in the actual calculation, but  for simplicity we neglect them in the schematic level diagram of Fig. \ref{fig4}.
In each panel we indicate the allowed transitions with arrows. Bilateral arrows indicate that the transition is allowed in both directions. The thickness of the arrows is roughly proportional to the transition probability.
Let us now explain the figure more in detail. For  $V\equiv V_1=156$ meV,
the potential equals the energy difference $E(N+1,3/2)-E(N,0)-V_G$. As can be seen in panel (c), the probability
of this transition is very small (the arrow is very thin). It is actually totally forbidden in the absence of spin-orbit coupling,
since the spin difference between the two states is $\Delta S=3/2$, i.e., it is larger than $1/2$.
In the presence of spin-orbit coupling, the small transition probability suffices to transfer population from the ground singlet, the only populated state for $V<V_1$, to the excited quartet $\vert 3/2 \rangle$. This yields 
a sudden change in the corresponding populations, visible in  panel (b) at $V=V_1$. In this new configuration, both transitions from $\vert 3/2 \rangle$ to $\vert 1\rangle$ and $\vert 2\rangle$ and back are possible, and their probability
is very high, as shown in Fig.~\ref{fig4}-c. This leads to a large sudden increase in the current, which can be seen
in panel (a). 
\begin{figure}[t]
	\centering
	\includegraphics[width=0.48\textwidth]{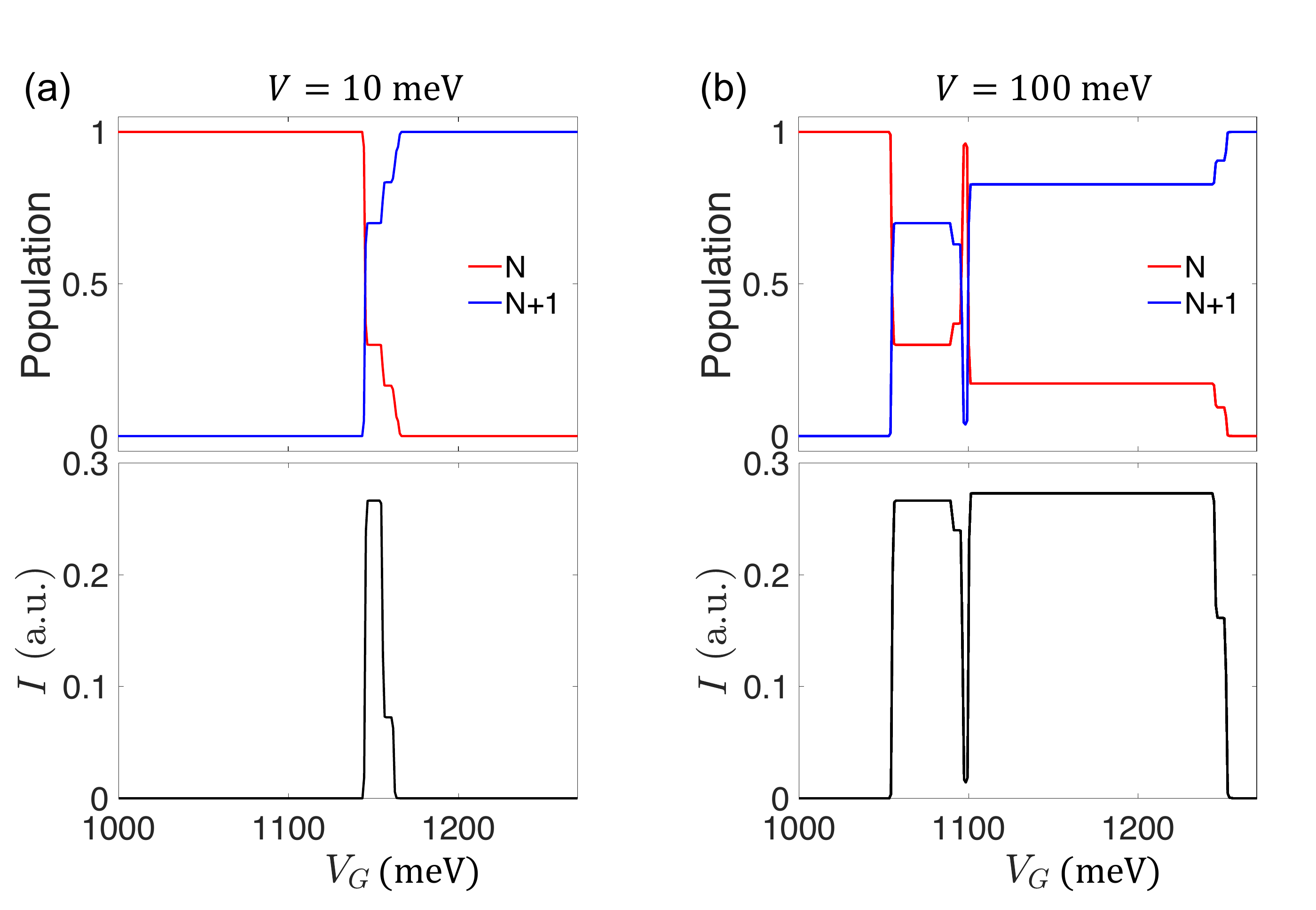} 
	\caption{(Color on-line) Population of the $N$ vs. $N+1$ states (top) and current (bottom) as a function of $V_G$, for two values of $V$ (panels a,b). }
	\label{figN}
\end{figure}
Further increasing the bias potential, one reaches the value $V=V_2=E(N+1,1/2)-E(N,2)-V_G$. 
As shown in panel (d), the (now in principle possible) transition
$|2\rangle \leftrightarrow |1/2\rangle$ has very small probability, since it
would be forbidden in the absence of spin-orbit interaction. On the other hand, as soon
as the $|1/2\rangle$ state is populated, the high-probability but unilateral transitions $|1/2\rangle \to |1\rangle$
and  $|1/2\rangle \to |0\rangle$ are possible. This leads in particular to a small jump  in the population $P_1$,
visible in panel (b). As a consequence, the current  slightly decreases, since some transport channels are blocked. 
This yields a negative conductance.
A similar phenomenon, with, however, a much stronger decrease in the current, occurs
at $V=V_3$. Here the  $\vert 1 \rangle \rightarrow \vert 1/2 \rangle$ transition becomes  accessible (see panel (e)
of Fig~\ref{fig4}), leading, via the unidirectional transition $|1/2\rangle\to|0\rangle$, to a large population transfer 
to the ground state of the neutral molecule, $|0\rangle$. 
In this situation the conductivity decreases to almost zero, as can be seen  in panel (a).
Due to the very small $\vert 0 \rangle \rightarrow \vert 3/2\rangle$ transition probability,
 the system remains almost locked in the $\vert 0 \rangle$ state 
till $V=V_4$.
Only when finally the high-probability $\vert 0 \rangle \rightarrow \vert 1/2\rangle$ transition is accessible, as shown in panel (f), the current flows again. This is shown in panel (a).\\

For completeness, in Fig.~\ref{figN}, we show similar results for fixed $V$ and as  a  function of $V_G$. 
The left  panel of the figure displays a weak bias case ($V=10$ meV). 
	The current sets on as far as the system remains within the bias window, i. e. for
	$\mu_R \le E(N+1,3/2)-E(N,0)-V_G \le \mu_L $.
	The figure shows that the population $P_{N+1}$ of the charged states increases monotonically with $V_G$. The first step in current and population  appears 
	for $V_G = E(N+1,3/2)-E(N,0)-V$;  the current is then suppressed for $V_G > E(N+1,3/2)-E(N,0)+V$. 
	Instead, the right panel of Fig.~\ref{figN} shows a case of larger bias ($V=100$ meV). Here the monotonic increase of $P_{N+1}$ is reversed  for $E(N+1,1/2)-E(N,2)-V \le V_G \le E(N+1,1/2)-E(N,0)-V$.
	This is exactly the same mechanism leading to negative differential conductance by varying $V$ at fixed $V_G$, illustrated earlier in the paper. The narrow region of negative differential conductance has a width corresponding to the splitting of the neutral molecular states, $E(N,2)-E(N,0)\approx 10$ meV. 

 \begin{figure}[t!]
	\centering
	\includegraphics[width=0.3\textwidth]{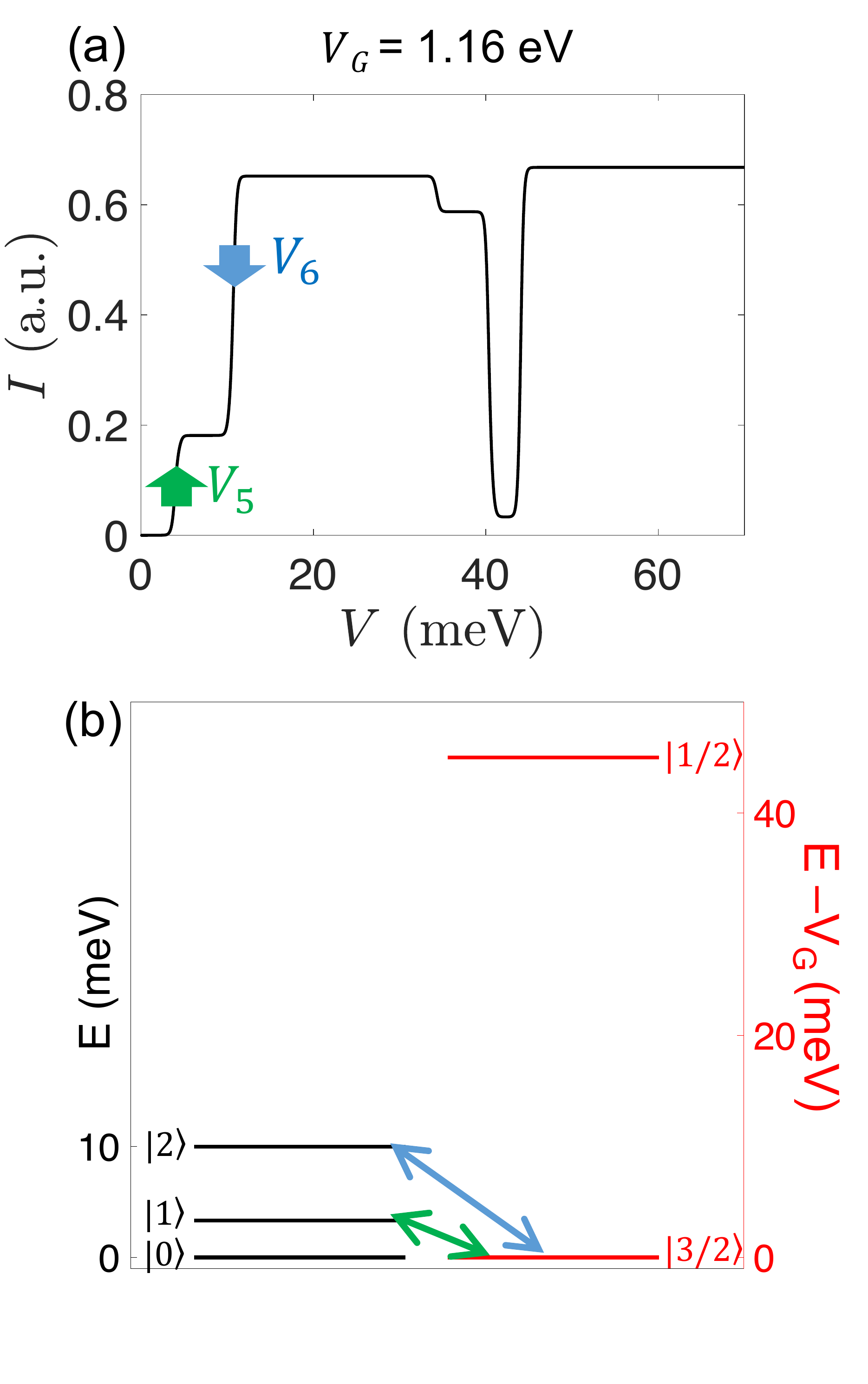} 
	\caption{(Color on-line) (a) Calculated $I(V,V_G)$ at $T=0$ for $V_G = 1.16$ eV, corresponding
		to the energy $E(N+1,3/2)-E(N,0)$. 
		The current is almost completely suppressed up to $V=V_5$, since the transition $|3/2\rangle\to |0\rangle$
		is forbidden in absence of spin-orbit coupling, and has very low probability otherwise. 
		Here $V_5=E(N+1,3/2)-E(N,1)-V_G\sim 3.3$~meV. A jump in conductivity is seen at  $V = V_6=E(N+1,3/2)-E(N,2)-V_G$.
		At this voltage the transition $|3/2\rangle\to|2\rangle$ is allowed (panel b).}
	\label{fig5}
\end{figure}

Going back to the case of fixed gate voltage and variable bias $V$,
Fig. \ref{fig5} shows similar effects than Fig.~\ref{fig4}, this time however for $V_G=E(N+1,3/2)-E(N,0)$.
In this case the current is suppressed for $V<V_5$, i.e., until the transition $\vert 3/2 \rangle \rightarrow \vert 1\rangle$ is accessible, since the $\vert 3/2 \rangle \rightarrow \vert 0 \rangle$  transition is forbidden in the absence of spin-orbit coupling, and has very weak probability otherwise.
This {\it spin-blockade} effect is a direct consequence of strong correlations which are explicitly included in our model.
A further increase in the current occurs at {$V=V_6$}, when the high-probability transition $\vert 3/2 \rangle \rightarrow \vert 2\rangle$ is accessible. 

Hence, we find for \{Ni$_2$\} spin-blockade effects and controllable regions of complete current suppression (on/off) in the stability diagram. 
This makes the system particularly interesting, since these phenomena enable one to electrically control the spin properties of the molecule, paving the way to potential spintronic and QIP applications.\cite{Chem} 
We stress that these features emerge as a consequence of  intra-molecular strong correlations, which we have explicitly included in the Hamiltonian (\ref{hubbard}). 
In particular, the Nagaoka mechanism discussed in the previous section plays a key role. 
While in bulk magnetic materials such a mechanism is of limited interest,  
the high degree of chemical control on the topology and strength of the exchange interactions makes single-molecule devices the ideal test-bed for Nagaoka-driven  phenomena.
Similar behaviors were evidenced with model-based effective models in other magnetic molecules, such as Mn$_{12}$ \cite{PRLWegewijs2006Mn12} and Co/Fe 2x2 grids,\cite{PRBWegewijs2007} where the addition/removal of a single electron to the neutral (half-filled) molecule changes the total spin from 0 to its maximum allowed value.
Also in these cases, this leads, in turn, to negative differential conductance and complete current suppression at finite bias voltages.\cite{PRBWegewijs2007} 
\begin{figure}[t]
	\centering
	\includegraphics[width=0.4\textwidth]{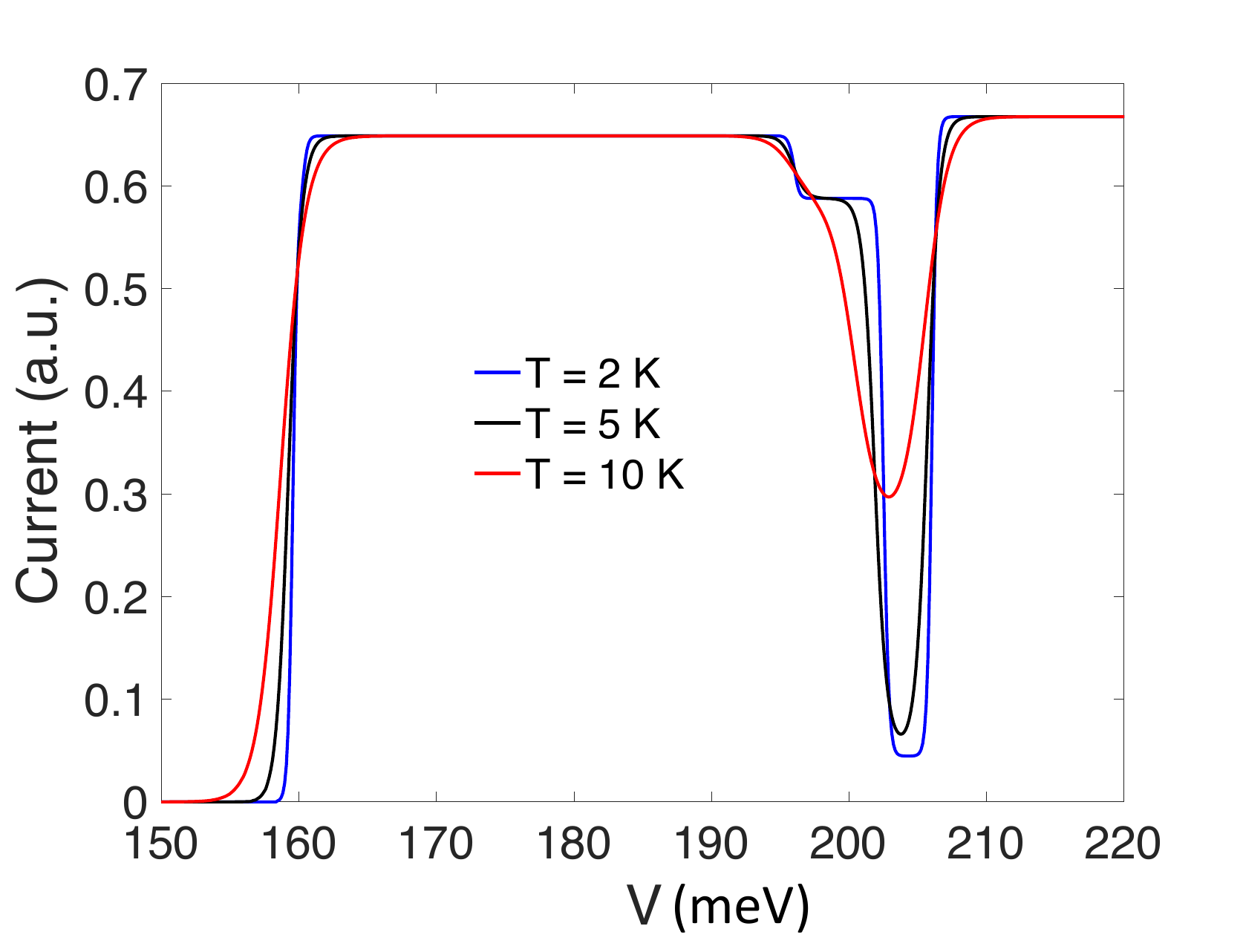} 
	\caption{(Color on-line) Effect of temperature on transport spectroscopy. The $I(V,V_G)$ curve is calculated for $V_G=1$ eV and as a function of the bias voltage $V$. By increasing the temperature, steps in the current are smoothed out, but the negative differential conductance region even for $T\sim10$ K. }
	\label{fig6}
\end{figure}

We would like to point out that, for the junction discussed here,  the step-structure of $I(V)$ is mostly determined by the specific form of the molecular many-body states and their energies, rather than by the relative strength of the tunnelling couplings in the hamiltonian $H_T$;
the latter, however, does modify the relative height of the steps. This can be shown by comparing our results with an idealized calculation in which the molecule-lead couplings are assumed to be the same for all states.

We now discuss the temperature dependence of the transport features discussed above. 
Fig. \ref{fig6} shows the $I(V, V_G)$ for $V_G=1$~eV (the same value used in Fig. \ref{fig4}),
this time calculated at different temperatures. The figure shows that, in the single-electron tunneling regime and for $k_BT \ll \mathcal{J}$, 
the equilibrium population of the molecular excited states is negligible and
the only effect of temperature is to smoothen out the steps in the  curve. It is worth noting that the negative differential conductance region is still present at $T=10$ K.

\begin{figure}[t!]
	\centering
	\includegraphics[width=0.4\textwidth]{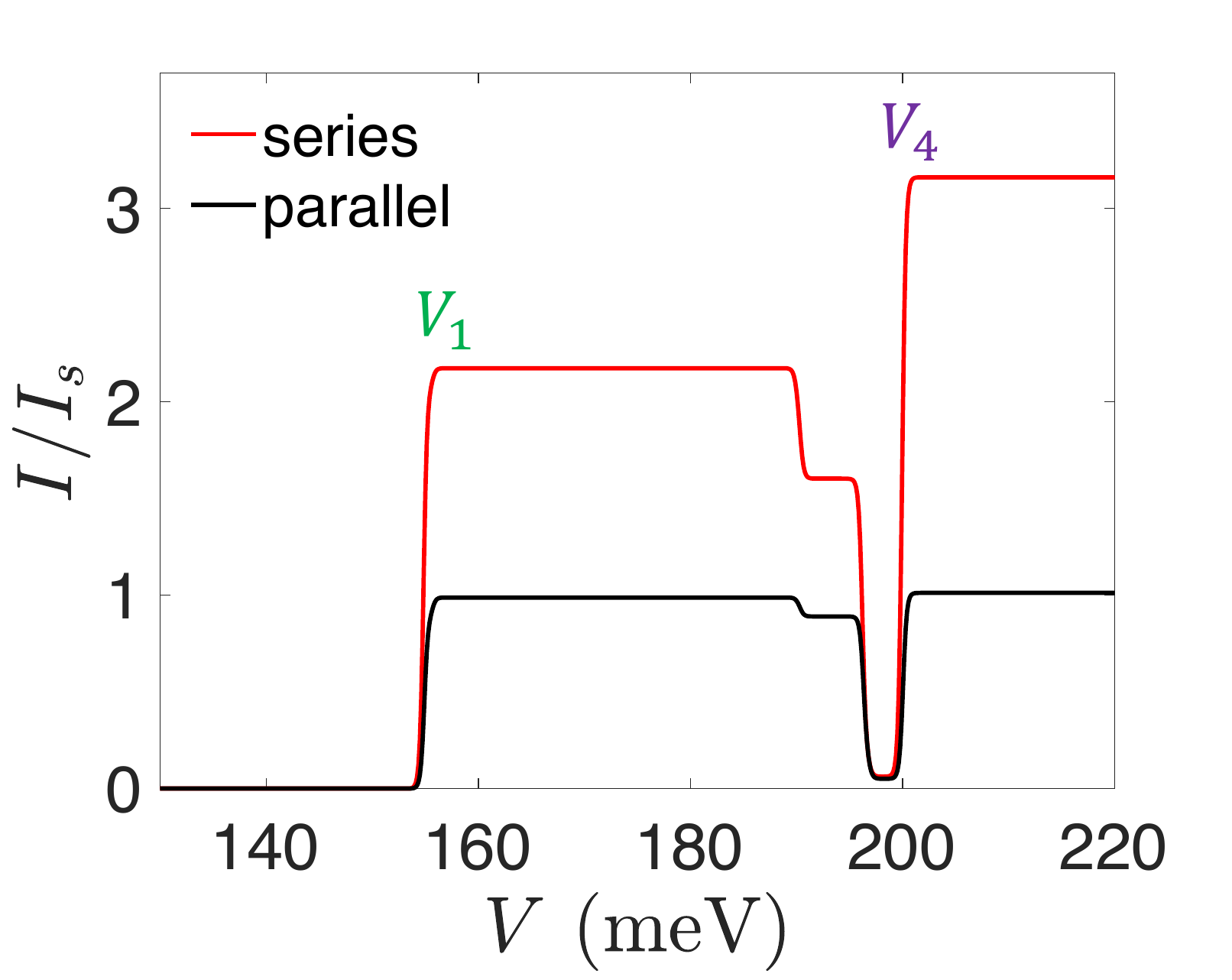} 
	\caption{(Color on-line) Serial versus parallel current for $V_G=1$~eV. This is the same value of the gate voltage used  in Fig.~\ref{fig4}. The current is normalized to its maximum value in the serial geometry. $V=V_1$: the current doubles by doubling the number of conducting channels. $V=V_4$: the current becomes three time as large, a quantum many-body effect related to the form of the molecular eigenstates and their specific
		coupling to the leads. 
		Here the potentials $V_1$ and $V_4$ are defined as in Fig.~\ref{fig4}.}
	\label{fig7}
\end{figure}

Finally, in  Fig.~\ref{fig7} we compare the transport properties of the serial configuration of the device, Fig. \ref{fig1}-(a), with those obtained for the parallel set up, Fig. \ref{fig1}-(b). In the latter each metal ion is connected to both electrodes.
By assuming equal tunneling rates for each conducting channel, switching from the serial to the parallel set up
increases the number of conducting channels from one to four; this leads to an enhancement of a factor four in the current.
The figure shows that, however, taking into account the actual changes in tunneling rates, the current only doubles for $V=V_1$,  and it becomes three times as large for $V=V_4$. 
This can be understood by analyzing the molecule-lead couplings $\gamma_{\lambda\lambda^\prime}^l$ in the two configurations. These involve a sum over orbitals and sites (see Eq. \ref{gamma}) which can give rise to partial sums or cancellations, depending on the specific structure of the molecule-lead hybridization for each pair of many-body states, $|\lambda\rangle$, $|\lambda^\prime\rangle$. In the serial case, ion 1 is only connected to the right lead, while ion 2 only to the right left. Therefore, the only relevant $\gamma_{\lambda\lambda^\prime}^l$ are
\begin{eqnarray} \nonumber
\gamma_{\lambda\lambda^\prime}^R &\propto& \sum_{\substack{m m^\prime\\ k \sigma}} \tau^{R1*}_{km} \tau^{R1}_{km^\prime} 
\langle \lambda_N^\prime \vert c_{1m\sigma} \vert \lambda_{N+1}  \rangle \langle \lambda_{N+1} \vert c^\dagger_{1 m^\prime \sigma} \vert \lambda_{N}^\prime \rangle, \\ \nonumber
\gamma_{\lambda\lambda^\prime}^L &\propto& \sum_{\substack{m m^\prime\\ k \sigma}} \tau^{L2*}_{km} \tau^{L2}_{km^\prime} 
\langle \lambda_N^\prime \vert c_{2m\sigma} \vert \lambda_{N+1}  \rangle \langle \lambda_{N+1} \vert c^\dagger_{2 m^\prime \sigma} \vert \lambda_{N}^\prime \rangle,
\end{eqnarray}
which are found to be always positive.
Conversely, in the parallel case, both ions are linked to both electrodes. Thus, there are contributions of the form
$$\gamma_{\lambda\lambda^\prime}^l \propto \sum_{\substack{m m^\prime \\ k \sigma}} \tau^{l1*}_{km} \tau^{l2}_{km^\prime} 
\langle \lambda_N^\prime \vert c_{1m\sigma} \vert \lambda_{N+1}  \rangle \langle \lambda_{N+1} \vert c^\dagger_{2 m^\prime \sigma} \vert \lambda_{N}^\prime \rangle,$$
some of which turn out to be negative. This yields a partial cancellation in the observed conductance and explains why it is reduced if compared to the naive picture in which the current merely increases linearly with the number of conducting channels.
Remarkably, we find that the actual enhancement can be tuned via changing the exact geometry
of the device, a property that could be used as a tool to optimize its performance.

\section{Conclusions}
In conclusion, we have introduced an efficient scheme to describe {\em ab-initio} quantum transport through molecular nanomagnets in the weak coupling regime.
This is an interesting regime for 
the electric control of molecular spin states, which can be used as a manipulation tool for spintronics or quantum information applications.\\
The approach is based on the DFT+MB method \cite{PRLdft} and treats both correlation effects and material aspects on the same footing.
We have applied this approach to a representative system, the \{Ni$_2$\} spin dimer. For this system we predict signatures of strong correlation effects such as spin-blockade, current suppression and negative differential conductance. 
We stress that such phenomena cannot be properly described within a mean-field description
of correlation effects, as adopted in methods based on simple approximations of the DFT exchange-correlation functional.
While the latter successfully describe the transport properties of weakly correlated systems,  
our method is suited for strongly correlated molecules. \\
These results
show the possibility of electronic control of the spin properties, making compounds like Ni$_2$ potentially very interesting for quantum information applications. 
For instance, one could exploit them as a switch of the interaction between a pair of molecular qubits.\cite{Chem,LossNatNano} By keeping the switch in the diamagnetic neutral state the effective qubit-qubit coupling is off, thus enabling the implementation of single-qubit rotations. Conversely, transition to the paramagnetic state of the anion can be exploited to activate an effective entangling evolution (e.g., XY or Heisenberg\cite{Chem,LossNatNano}) within the two-qubits computational subspace. 
In order to asses the actual feasibility of the proposed quantum computational schemes, our calculation could be extended to include a pair of qubits linked through the \{Ni$_2$\} switch.  \\

\textit{Acknowledgements}.
The authors acknowledge financial
support from the Italian Ministry of Education and Research (MIUR) through PRIN Project 2015 HYFSRT  ``Quantum Coherence in Nanostructures of Molecular Spin Qubits'', from the European Project SUMO of the call
QuantERA and from the Deutsche Forschungsgemeinschaft through the research training group RTG1995.
A.C. also acknowledges ``Fondazione Angelo Della Riccia" for supporting this project. 
Calculations were done on the J\"ulich supercomputer JURECA and JUWELS.


\begin{thebibliography}{99}
\bibitem{geometry}
Since our main goal is to present an efficient scheme for the description of transport properties through strongly-correlated molecules, here we do not perform an optimization of the junction geometry. If necessary, this can be done with standard methods once the details of the actual experimental implementation are known. In the weak coupling regime, however, the effects of such geometry optimization are expected to be small.
\end{thebibliography}
\end{document}